\documentclass[twocolumn,fleqn,usenatbib]{mnras}

\usepackage{mathptmx}

\usepackage[T1]{fontenc}
\usepackage{ae,aecompl}

\usepackage{graphicx}	% Including figure files
\usepackage{amsmath}	% Advanced maths commands
\usepackage{amssymb}	% Extra maths symbols
\usepackage{breqn}

\newcommand{\diff}{\mathrm{d}} % total derivative shortcut
\newcommand{\romane}{\mathrm{e}} % total derivative shortcut

\newcommand{\diffrac}[2]{ \frac{\diff {#1}}{\diff {#2}} } % total derivative shortcut
\newcommand{\parfrac}[2]{ \frac{\partial {#1}}{\partial {#2}} } % partial derivative

\newcommand{\parsub}[1]{ \partial_{#1} } % partial derivative

\newcommand{\bruntvaisala}{Brunt-V\"{a}is\"{a}l\"{a} }

\newcommand{\unitg}{cm~s$^{-2}$} % unit of acceleration
\newcommand{\unity}{g~cm$^{-2}$} % unit of column depth
\newcommand{\unitF}{ergs~cm$^{-2}$~s$^{-1}$} % unit of heat flux

\newcommand{\Msol}{$\mathrm{M}_{\odot}$}
\newcommand{\mEdd}{\dot{m}_\mathrm{Edd}}
\newcommand{\MeVnuc}{MeV~nuc$^{-1}$}

\newcommand{\Vcorr}{\mathcal{V}}

\title[Relativistic ocean $r$-modes]{Relativistic ocean $r$-modes during type-I X-ray bursts}

\author[F. R. N. Chambers and A. L. Watts]{
F. R. N. Chambers,$^{1}$\thanks{E-mail: frnchambers@uva.nl}
A. L. Watts$^{1}$
\\
$^{1}$Anton Pannekoek Institute for Astronomy, University of Amsterdam, Postbus 94249, 1090 GE Amsterdam, The Netherlands\\
}

\date{Accepted XXX. Received YYY; in original form ZZZ}

\pubyear{2019}

\begin{document}
\label{firstpage}
\pagerange{\pageref{firstpage}--\pageref{lastpage}}
\maketitle

\begin{abstract}
Accreting neutron stars (NS) can exhibit high frequency modulations in their lightcurves during thermonuclear X-ray bursts, known as burst oscillations. These frequencies can be offset from the NS spin frequency by several Hz (where known independently) and can drift by $1-3$ Hz. One plausible explanation is that a wave is present in the bursting ocean, the rotating frame frequency of which is the offset. The frequency of the wave should decrease (in the rotating frame) as the burst cools hence explaining the drift. A strong candidate is a buoyant $r$-mode. 
To date, models that calculated the frequency of this mode taking into account the radial structure neglected relativistic effects and predicted rotating frame frequencies of $\sim 4$~Hz and frequency drifts of $> 5$~Hz; too large to be consistent with observations. We present a calculation that includes frame-dragging and gravitational redshift that reduces the rotating frame frequency by up to $30 \%$ and frequency drift by up to $20 \%$. 
Updating previous models for the ocean cooling in the aftermath of the burst to a model more representative of detailed calculations of thermonuclear X-ray bursts reduces the frequency of the mode still further. This model, combined with relativistic effects, can reduce the rotating frequency of the mode to $\sim 2$~Hz and frequency drift to $\sim 2$~Hz, which is closer to the observed values.
\end{abstract}


%%%%%%%%%%%%%%%%%%%%%%%%%%%%%%%%%%%%%%%%%%%%%%%%%%
%%%%%%%%%%%%%%%%% BODY OF PAPER %%%%%%%%%%%%%%%%%%

\section{Introduction}

Type I X-ray bursts are caused by runaway thermonuclear burning in the ocean layer of a neutron star (NS). This layer is composed of matter accreted from the low-mass companion star and ashes left over from previous bursts. These bursts reveal information about the nuclear processes ongoing in the ocean of the star which can vary depending on fuel, temperature, and accretion rate \citep[][]{Lewin93,Bildsten98b,Strohmayer06,Galloway17}. This paper focuses on the phenomenon of \textit{thermonuclear burst oscillations} (TBOs). Timing analysis of some type I X-ray bursts reveals periodic oscillations in the lightcurve. These oscillations are thought to arise due to asymmetries in surface brightness \citep{Strohmayer97a} but the mechanism responsible remains unknown. The observed frequencies are either at the spin frequency of the NS, or offset by a few Hz and may drift by several Hz during the tail of the burst (for an overview of type I bursts and TBOs see, \citealt{Galloway08} and \citealt{Watts12}).

Various mechanisms have been suggested to explain the origin of these asymmetries. These include the flame front spreading across the surface of the star \citep{Spitkovsky02,Cavecchi13,Cavecchi15,Cavecchi16}; and a cooling wake model examining the idea that different areas of the star cool at different rates \citep{Mahmoodifar16}. However, these models have some issues: the flame spread model can account for oscillations in the rise but not the tail\footnote{Recent studies by \cite{Cavecchi19} indicate that large scale vortices may be set up in the aftermath of the burning phase of the burst; these may give rise to fluctuations in the tail of the lightcurve.}; and the cooling wake model requires enhanced cooling from an as yet unidentified mechanism. It is also possible that convection in the ocean needs to be considered \citep{Medin15,Garcia18a,Garcia18b}, or that the magnetic field of the star plays a key role in explaining TBOs \citep{Heng09}. More extensive studies of these possibilities are required.

This paper will explore the idea that a global mode in the ocean of the star is excited during the burst, giving rise to large-scale patterns. The frequency of the mode in the inertial frame would be the TBO frequency, while the frequency in the rotating frame would be the offset from the NS spin. Sources which exhibit drift in the tail of the burst are naturally explained by a mode model; the frequency of the mode would depend on the condition of the ocean, and hence drift as the ocean cools\footnote{Note that TBOs observed from sources that show persistent accretion-powered pulsations are not well-explained by this model since they show frequency drift predominantly in the rising phase of the burst \citep{Chakrabarty03,Strohmayer03,Altamirano10a,Cavecchi11}.}. \cite{Heyl04} suggested the global mode model as an explanation for TBOs and cut down the many different families of potential modes based on some key observations: the TBO frequency is close to the spin frequency of the NS, so the rotating frame frequency of the mode should be $\sim 1$ Hz and the azimuthal wavenumber $m=1$; since most observed TBOs drift upwards towards the spin frequency as the ocean cools\footnote{Although some bursts show a TBO whose frequency decreases in the tail of the burst; for example, a few bursts from 4U~1728-34 and 4U~1636-536
\citep{Muno02a}.}, the modes should travel in the opposite sense to the star's rotation (retrograde); the modes should maximise visibility (pulsed amplitude) so a mode with a small number of latitudinal nodes will be preferred. Given these constraints, the best candidate is a low azimuthal wavenumber buoyant $r$-mode, which is driven by buoyancy in the ocean and strongly affected by the Coriolis force.

One problem emerging from the initial calculations by \citet{Heyl04} of the buoyant $r$-mode model was an over-prediction of the frequency drift at $> 5$ Hz. \citet{Lee04} calculated the radial structure of these modes for two ocean models representing the early and late stages of the burst in order to estimate frequency drift, and found that the drifts were too large to be consistent with observations. Further work by \cite{Piro05b} (hereafter PB05) calculated frequency drift directly by including a model for the ocean cooling in the aftermath of the burst. Snapshots of the ocean from this cooling model were used as a background upon which to calculate the mode frequency. They also found drifts too large ($> 5$ Hz) to be consistent with observations, prompting the suggestion that the ocean mode might transition into a crustal interface wave \citep{Piro05a} during the tail of the burst, thereby halting the drift. This particular idea was ruled out by \citet{Berkhout08}, due to the weak coupling between the ocean mode and the crustal interface wave.

Since the work of PB05, a wealth of new models of thermonuclear X-ray bursts have been developed and fitted to bursts from specific sources; these models include a full treatment of the nuclear reactions in the ocean \citep{Heger07b,Keek17,Meisel18a,Johnston18,Johnston19}. A new heat source, \textit{shallow heating}, has been suggested to resolve inconsistencies between observations and the theory needed to explain crust cooling models \citep[for a review see][]{Wijnands17} and has implications for other phenomena such as short waiting time bursts, superburst recurrence times, and the transition between different burst regimes \citep{Gupta07,Keek11,Keek17}. Including these effects can have a marked affect on frequency drift; \cite{Chambers19} calculated buoyant $r$-modes in a bursting ocean model developed by \cite{Keek17} that included both a changing composition and enhanced heat flux from the crust. Frequency drift of the TBOs was reduced to $< 4$ Hz, but the rotating frame frequency increased by $2-4$~Hz. This work only inspected one burst model, however, and since oscillations appear in a variety of bursts from different sources, a wider variety of bursts need to be tested in order to draw more general conclusions about the mode frequency.

The outer layers of a NS are a relativistic environment where effects such as gravitational redshift and frame-dragging play an important role. \citet{Maniopoulou04} (hereafter MA04) derived a set of perturbation equations which include gravitational redshift and frame-dragging, for a mode in a shallow layer on the surface of a relativistic star. MA04 estimated the magnitude of the influence of these relativistic effects on the frequencies of various classes of modes, finding a decrease in rotating frame frequency of $15-20\%$ as compared to a Newtonian calculation. However, this approach assumed plane wave solutions for the radial structure of the mode, and did not model the cooling ocean in order to calculate frequency drift.

In this work we take this next step, performing a full calculation of the radial profile and frequency of buoyant $r$-modes including relativistic effects, upon a NS ocean cooling in the aftermath of a burst. We inspect how the conditions in the ocean influence the buoyant $r$-mode using a simplified model for ocean cooling, in order to estimate how frequencies will change when considering more realistic models of burst conditions.

\section{Relativistic mode equations}
\label{sec:perterbation-equations}

We make use of the set of perturbation equations developed by MA04: in what follows we summarise the key points from that paper, referring the reader to the original publication for more details.

MA04 use the background state of the fluid and space-time of a slowly rotating star first given in \citet{Hartle67}, which includes gravitational redshift and the dragging of inertial frames. This metric is valid for stars rotating at $\lesssim 800$~Hz \citep[for a more detailed discussion of the validity of the Hartle-Thorne metric see][]{Berti05,Baubock13} and MA04 keep only terms linear in the star's rotation and frame-dragging meaning that rotation rates should, in principle, be somewhat smaller than this for these equations to be applicable. Since the highest frequency TBO observed is $620$~Hz for 4U~1608-522 \citep{Hartman03}, we assume that these approximations are still valid. Deformation of the star from a spherical shape is an extra effect at quadratic order in the rotation rate and not taken into account in this calculation. Assuming the star to be made of a perfect fluid, the Einstein equations and hydrostatic equilibrium reduce to the Tolman-Oppenheimer-Volkoff (TOV) equations (stellar structure and space-time for a non-rotating star) and an extra differential equation for frame-dragging. 

The mode equations are derived by solving for perturbations upon this background. The Cowling approximation is assumed, which fixes the space-time to be constant, leaving only the velocity, pressure and mass-energy of the fluid to be perturbed. The perturbations depend sinusoidally on time $t$ and azimuth $\phi$ as $\exp \left( im \phi + i \sigma t \right)$, with an inertial frame frequency $\sigma$ and azimuthal wavenumber $m$. The \textit{traditional approximation}, a standard approximation in the geophysical literature \citep{Eckart60,Pedlosky87}, is made; this approximation assumes that the Coriolis force in the radial direction is negligible compared to the other forces (such as buoyancy) and that the fluid motion is mostly horizontal, not vertical. Type I X-ray bursts occur in a shallow layer $\lesssim 10^4$~cm thick on the surface of a NS. Since the radius of the star is $\sim 10$~km, the traditional approximation is valid (see MA04 for further details). The perturbations are assumed to be adiabatic, which relates perturbations in pressure to perturbations in mass-energy of the fluid.

\subsection{Specialising to the neutron star ocean}
\label{subsec:thin-layer-approx}

MA04 give a set of two differential equations (equations 53 and 54 of that paper\footnote{Note that in MA04, there is a factor of $(p+\rho)$ missing in the first term of equation 53; this is corrected later in the paper.}) to solve for $\delta p$, the Eulerian perturbation of the pressure, and $\xi_r$, the Lagrangian fluid displacement in an orthonormal basis. In these equations appear the potentials $\nu(r)$, $\lambda(r)$ and $\omega(r)$ that describe gravitational redshift and frame-dragging. In our analysis these potentials are approximated to their value on the boundary of the star (at $r=R$). Restoring units, these become:
\begin{subequations}
  \begin{align}
    &\romane^{- \nu  / 2} \approx \romane^{\lambda / 2} 
    \approx \left( 1 - \frac{2 G M}{c^2 R} \right)^{-1 /  2} \equiv \Vcorr , \\
    & \frac{1}{2} \diffrac{\nu}{r} \approx - \frac{1}{2} \diffrac{\lambda}{r} \approx \frac{g \Vcorr}{c^2} , \\
    &\omega \approx \frac{2 \Omega I G }{c^2 R^3} ,
  \end{align}
\end{subequations}
where $\Vcorr$ is the volume correction factor, $g = G M \Vcorr / R^2$ the local gravitation acceleration, and $I$ the moment of inertia of the NS.
With these approximations, the hydrostatic equilibrium condition is $\diff p / \diff r = - \rho g \Vcorr$, and the two perturbation equations from MA04 become:
\begin{subequations}
  \begin{dmath}
    \label{eq:dp-layer}
    \parfrac{}{r} \frac{\delta p}{p}
    =
    \frac{\rho \Vcorr}{p} \left[ \left(\sigma + m\Omega\right)^2 \Vcorr^2 + g \mathcal{A} \right] \xi_{r}
    + \frac{g \rho \Vcorr}{p} \left( 1 - \frac{1}{\Gamma_1} \right) \frac{\delta p}{p} ,
  \end{dmath}
  \begin{dmath}
    \label{eq:xi-layer}
    \parfrac{\xi_{r}}{r}
    =
    \xi_{r} \left( \frac{\rho g \Vcorr}{p \Gamma_1} - \frac{2}{r} \right)
    +
    \left( \frac{L_\mu}{ \left(\sigma + m\Omega\right)^2 r^2} \frac{p}{\rho \Vcorr} - \frac{\Vcorr}{\Gamma_1} \right) \frac{\delta p}{p}
  \end{dmath}
\end{subequations}
where $L_\mu$ is an operator that will be discussed in Section \ref{subsec:laplace-equn}, $p$ and $\rho$ are the fluid pressure and mass density, $\xi_r$ is the radial component of the Lagrangian fluid displacement in an orthonormal basis, $\Gamma_1$ is the first adiabatic index, and
\begin{equation}
    \label{eq:rel-sch}
    \mathcal{A} = \Vcorr^{-1} \left( \frac{1}{\rho} \diffrac{\rho}{r} - \frac{1}{\Gamma_1 p} \diffrac{p}{r} \right) ,
\end{equation}
is the relativistic Schwarzschild discriminant. These equations reduce to the Newtonian limit when $\Vcorr = 1$ and $\omega = 0$. The inertial frame frequency of the mode, $\sigma$, only appears in the equations combined with the stellar rotation frequency, $\Omega$. For convenience we refer to this term as the rotating frame frequency, $\sigma_r \equiv \sigma + m\Omega$, even though the true rotating frame frequency should include a factor of $\Vcorr$ due to gravitational redshift.

\subsection{Solutions to Laplace's Tidal Equation}
\label{subsec:laplace-equn}

The operator in Equation (\ref{eq:xi-layer}) appears in Laplace's Tidal Equation: $L_\mu f = - \lambda f$, $\lambda$ being the eigenvalue. $L_\mu$ depends on the co-latitude through a convenient coordinate $\mu \equiv \cos\theta$, and the spin parameter $q$ as:
\begin{dmath}
  \label{eq:laptide}
  L_\mu f \equiv
  \diffrac{}{\mu} \left( \frac{1 - \mu^2}{1 - q^2 \mu^2} \diffrac{f}{\mu} \right)
  - q m \frac{ 1 + q^2 \mu^2}{\left( 1 - q^2 \mu^2 \right)^2} f
  - \frac{m^2}{\left(1 - \mu^2\right) \left( 1 - q^2 \mu^2 \right)} f .
\end{dmath}
In the Newtonian calculation, the spin parameter only depends on the frequency of the mode and the spin frequency of the star, but including frame-dragging alters the definition of the spin parameter to:
\begin{equation}
  \label{eq:spin-param}
  q \equiv \frac{ 2 \Omega }{\sigma_r} \left( 1 - \frac{\omega}{\Omega} \right) = q_{\mathrm{N}} \left( 1 - \frac{\omega}{\Omega} \right) ,
\end{equation}
where setting frame-dragging to zero, $\omega = 0$, returns the Newtonian expression, $q_{\mathrm{N}}$.

The solutions of Laplace's Tidal Equation are discussed in many works \citep[see][]{Longuet68,Lee97,Townsend03}. As we are interested in a single buoyant $r$-mode as the mechanism responsible for TBOs, we restrict $\lambda$ to conform to the $r$-mode family of solutions. The $r$-modes represent a special case for Laplace's Tidal Equation, since if the value of the spin parameter is greater than a certain threshold (for example, $q \gtrsim 10$ for the $m=1$, $l=2$ mode which is satisfied by the NSs studied here, see Section \ref{subsec:surface-pattern}) then the eigenvalue, $\lambda$, becomes independent of the spin parameter (see figure 4 from PB05)\footnote{This does not mean that the eigenfunction $f(\mu = \cos \theta)$ is independent of the spin parameter. This will be discussed in Section \ref{subsec:surface-pattern}.}. For a spin parameter above this threshold, the term $L_\mu \delta p$ in Equation (\ref{eq:xi-layer}) can be replaced with $- \lambda \delta p$, where a value for $\lambda$ appropriate for an $r$-mode is chosen. This procedure cannot be performed with rotationally modified $g$-modes as they have a quadratic dependence on the spin parameter.
The most promising candidate to explain TBOs identified by \citet{Heyl04} and PB05 was the $m=1, l=2$ buoyant $r$-mode, as the frequency would be close to that of the NS spin frequency and result in the highest variability in the angular component. However, other authors have considered higher latitudinal number modes ($l=3, 4$ in \citealt{Lee04} and $l=3$ in \citealt{Lee05}). We will also consider these modes, and will show that they result in a smaller rotating frame frequency. The values of $\lambda$ for the $m = 1$ buoyant $r$-mode are $1.1 \times 10^{-1}$, $4.1 \times 10^{-2}$, and $2.2 \times 10^{-2}$ for $l=2$, $l=3$, and $l=4$ respectively. 
\footnote{The $m=l=1$ mode is the mixed gravity-Rossby wave, sometimes referred to as a Yanai mode \citep[see][for further details]{Townsend03}. For rapid rotation, this mode becomes highly equatorially trapped in a similar fashion to rotationally modified $g$-modes and so is not included in this work.}

\subsection{Method of solution}

In Section \ref{sec:cooling} we describe our model for how the background state of the NS ocean evolves in the aftermath of a burst. This model provides a set of values for $p(r)$, $\rho(r)$, $\Gamma_1(r)$ and $\mathcal{A}(r)$ in the ocean for a series of snapshots in time. We solve the perturbation Equations (\ref{eq:dp-layer},\ref{eq:xi-layer}) upon this background.

A shooting method is used to search for the rotating frame frequency with the outer boundary conditions that the Lagrangian pressure perturbation must be zero (which provides a relation between $\delta p$ and $\xi_{r}$) and that $\xi_r$ must be zero at the crust. The outer boundary for the mode is chosen to be at a depth where the thermal timescale and mode timescale are approximately equal (at $\sim 10^7$~\unity~in column depth, defined in Section \ref{subsec:cooling-equns}). The mode is somewhat sensitive to the location of this boundary, but we expect the condition to be robust since the energy of the mode is concentrated at least an order of magnitude deeper than $10^7$~\unity. See Section \ref{subsec:result-hp} for further discussion. The mode frequency is not sensitive to the precise location of the inner boundary at the crust; we tested this by varying the crust column depth, as in \citet{Chambers18}.

There are several effects to consider when choosing a normalisation for the mode. When inspecting the effect of this mode on the lightcurve from the star during a burst, normalisation will become very important as it will dictate the amplitude across the surface. Observations would place a constraint on the maximum perturbation in flux, which could be related to temperature and the pressure perturbations \citep{Piro06}. Coupling of the mode to the photosphere of the NS may play a role here. Another choice might be to consider the maximum allowable temperature perturbation in the ocean that would not rekindle another burst. PB05 used an energy condition to normalise modes. \citet{Lee05} normalised these modes based on the toroidal component of the mode displacement Since the normalisation has no effect on the frequency, we use the simple condition that $\delta p / p = 1$ at the surface.

\section{Cooling ocean model}
\label{sec:cooling} 

We have chosen to examine simplified models for the cooling ocean in the aftermath of a thermonuclear X-ray burst, in order to facilitate the most direct comparison with the previous calculations of PB05. We test a new parametrisation of the initial condition which better approximates the temperature and density profile of the ocean in more detailed calculations of thermonuclear X-ray bursts (such as those carried out by \citealt{Keek17,Johnston18,Johnston19}) by removing the large discontinuity in density present in the model of PB05. This new model will not take into account nuclear burning ongoing throughout the burst, which influences the frequency drift exhibited by the buoyant $r$-mode \citep{Chambers19}. The composition of the new model is chosen to match the calculations of PB05, not \citet{Keek11}. This will mainly affect the rotating frame frequency; for a discussion see Section \ref{subsec:dis-nucback}. There are two key differences in the new model for the initial conditions: the first is to distribute the energy released by the burst in such a way as to remove large discontinuities in density close to the ignition site; and the second is to increase the flux coming from the crust by a factor of ten, which explores the effect of shallow heating (not included in PB05's model).

\subsection{Equations for temperature evolution}
\label{subsec:cooling-equns}

In order to follow the thermal evolution of the ocean in the tail of the burst, we use relativistic equations from \citet{Brown09} that describe the evolution of a star in quasi hydrostatic equilibrium. We do not expect rotation to have a strong effect on this portion of the calculation since we are examining a slowly rotating background. All the fuel for unstable thermonuclear burning is assumed to be used up during the initial explosion, and radiative cooling in the aftermath of this explosion is expected to be the primary source of temperature change. Sources of energy from ongoing nuclear burning are therefore neglected. This is also in keeping with the assumptions of PB05. We apply the same approximations described in Section \ref{subsec:thin-layer-approx} to transform the equations for temperature evolution to a form appropriate for the shallow ocean on a NS. For temperature $T$, and heat flux $F_{r}$ in the radial direction, the temperature evolution follows:
\begin{subequations}
  \begin{align}
    \label{eq:T-evol}
    c_p \parsub{t}{T} &= - \frac{\Vcorr^{-2}}{\rho r^2} \parfrac{}{r} \left( r^2 F_r \right) , \\
    \label{eq:flux}
    F_r &= - \Vcorr^{-1} K \parfrac{}{r} \left( T \right) ,
  \end{align}
\end{subequations}
where $c_p$ is the heat capacity at constant pressure and $K$ is the conductivity. Both of these quantities depend on the state of the fluid.

Using the hydrostatic equilibrium condition, derivatives with respect to radius are changed to a more convenient variable, column depth; the mass density of fluid above a point within a column beginning at the surface of the star. For this work we define it as $\diff y \equiv - \Vcorr \rho \diff r$, which means that the equations describing the temperature evolution of the ocean in the aftermath of the burst are:
\begin{subequations}
  \begin{align}
    \label{eq:T-evol-NS-layer}
    c_p \parfrac{T}{t} &= \Vcorr^{-1} \parfrac{F_r}{y} + \Vcorr^{-2} \frac{2 F_r}{\rho R} , \\
    \label{eq:flux-NS-layer}
    F_r &= \rho K \parfrac{T}{y}  , \\
    \label{eq:hydrostat-NS-layer}
    p &= y g  , \\
    \label{eq:radius-NS-layer}
    \Vcorr r &= \Vcorr R - \int_0^y \frac{1}{\rho} \diff y  ,
  \end{align}
\end{subequations}
which reduces to the Newtonian calculation for $\Vcorr = 1$.

\subsection{Equation of state and opacity}

The ocean consists of a fully ionised electron-ion plasma in varying degrees of degeneracy. The equation of state, opacity, and conductivity of the fluid are calculated as described in \citet{Chambers18}. The equation of state is that of \citet{Potekhin10}. The radiative opacity, which is dominant in the shallowest portion of the ocean, contains contributions from free-free and electron scattering \citep{Schatz99} which are combined using a \textit{non-additivity} factor \citep{Potekhin01}. Conductivity dominates in the more dense region of the ocean and is mostly mediated by electrons \citep{Potekhin99}.

\subsection{Initial conditions}
\label{subsec:init-cond}

\begin{figure}
  \centering
  \input{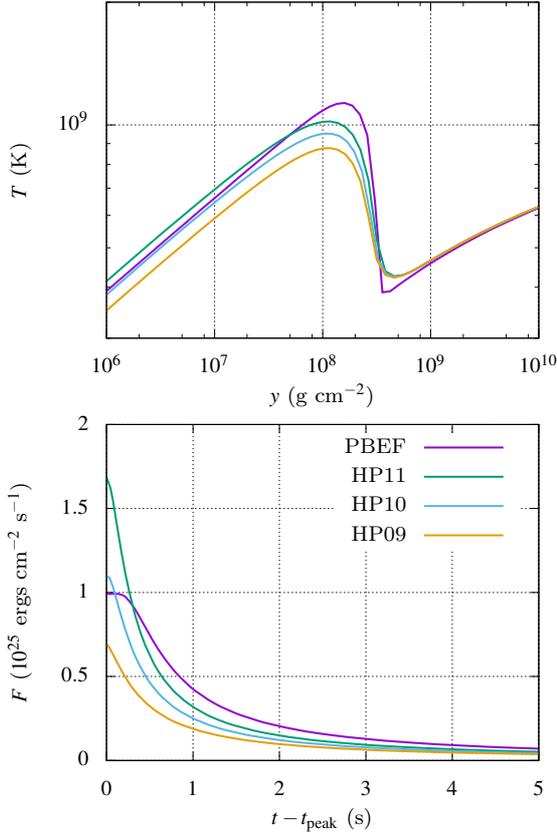}
  \caption{
    The temperature profile $0.1$~s after the peak of the burst, and the flux at the outer boundary for several models of the cooling ocean in the aftermath of a burst. All of these calculations assumed a NS of mass $1.4$~\Msol~and radius $10$~km, and include relativistic effects. The purple lines are from a model that uses the initial conditions defined in PB05 with a flux emanating from the crust of $F_{\mathrm{crust}} = 10^{22}$~\unitF. The green, blue and orange lines are for the heat pulse model suggested in this work, with peak temperatures $1.1$, $1.0$, and $0.9$~GK respectively (models HP11, HP10, and HP09). Buoyant $r$-modes are calculated on these cooling backgrounds from 0.1~s after the peak of the burst.}
  \label{fig:heat-pulse-background}
\end{figure}

\begin{figure}
  \centering
  \input{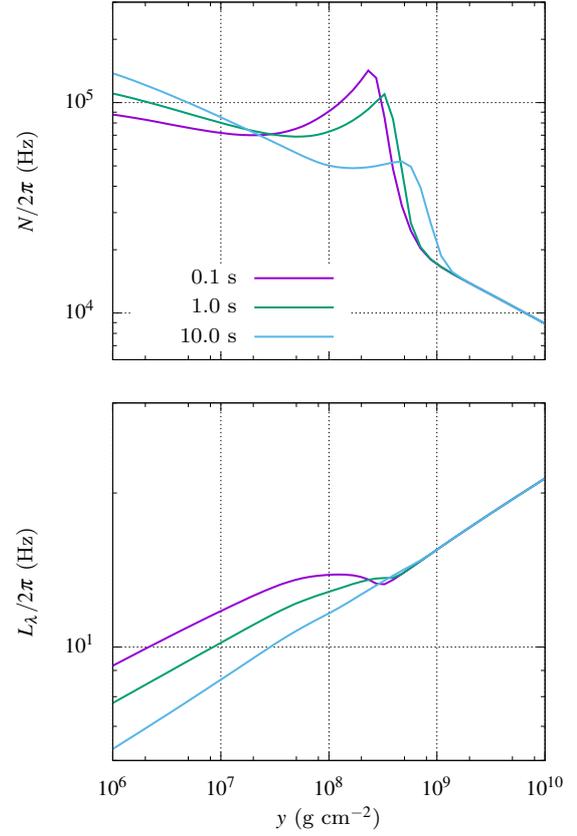}
  \caption{The \bruntvaisala frequency and the Lamb frequency for the cooling model HP10 at the times 0.1~s, 1.0~s, and 10~s after the peak of the burst. The top panel shows the \bruntvaisala frequency which exhibits a bump at approximately $y = 3 \times 10^{8}$~\unity{} as a result of the gradient in density and temperature between the hot and cold layers of the ocean. The Lamb frequency, shown in the lower panel, is calculated from the formula $L_\lambda = \sqrt{\lambda} c_s / R$, where $c_s$ is the sound speed and $\lambda = 0.11$ appropriate for an $m=1$, $l=2$ buoyant $r$-mode. In the deep ocean both frequencies are well described by a totally degenerate, ultra-relativistic gas of electrons and obey the scaling laws $N \propto \rho^{-1/3} \propto p^{-1/4}$, and $L_\alpha \propto \rho^{1/6} \propto p^{1/8}$.}
  \label{fig:heat-pulse-bv-lamb}
\end{figure}

The initial temperature profile used by PB05 divides the NS ocean into two layers, a bursting layer and a cool layer, each of which is assumed to have constant flux. The bursting layer, in which heat from nuclear burning is deposited, assumes a heat flux of $10^{25}$~\unitF. The cool, deep layer is set by the flux emanating from the crust, which is assumed to be $10^{21}$~\unitF. The composition is fixed in both layers. The bursting layer is assumed to be pure $^{40}$Ca to represent post-burst ashes and the cool layer pure $^{64}$Zn \citep[both inspired by][]{Schatz01,Woosley04}. There is a discontinuity in temperature and density at the transition between these two layers, the location of which is set by the conditions for thermonuclear runaway at a column depth of $3 \times 10^{8}$~\unity{} \citep[see][]{Bildsten98b}. A NS with radius $10$ km and mass $1.4$~\Msol~was assumed, which results in a gravitational acceleration of $1.9 \times 10^{14}$~\unitg{} (relativistic effects are not included in PB05, therefore $\Vcorr = 1$).

In this work, three versions of PB05's model will be tested in order to demonstrate the influence of gravitational redshift and frame-dragging. The first model uses the same parameters as the original paper and does not include relativistic effects. As such, this model will be referred to as PB05. The second and third versions include relativistic effects and use the same value of the gravitational acceleration as PB05, changing one of either the mass or the radius to do so, since $\Vcorr$ is no longer zero \citep[see][for more details]{Keek11}. The model with the same radius or mass as PB05 will be referred to as GRR or GRM respectively (see Table \ref{table:PB05-params} for a complete list of values).

\begin{table}
  \centering
  \caption{The parameters that outline each cooling model used to test relativistic effects on the $m=1$, $l=2$ buoyant $r$-mode. $\Vcorr = 1$ indicates relativistic effects are not included.
  }
  \label{table:PB05-params}
  \begin{tabular}{l r r r}
  Model & $M$ (\Msol) & $R$ (km) & $\Vcorr$ \\
    \hline
    PB05  & 1.4  & 10     & 1    \\ 
    GRR   & 1.14  & 10    & 1.23 \\ 
    GRM   & 1.4   & 11.2  & 1.26 \\ 
  \end{tabular}
\end{table}

Our new model for the initial condition is the sum of a pre-burst state, that holds flux constant throughout the layer, and a heat pulse, representing the energy released by nuclear burning. This new model provides a `rising phase' to the burst where heat spreads rapidly from the centre of the pulse to the shallow ocean until there is a peak in flux at the outer boundary; time is set to zero at this point. The key difference between PB05's initial temperature profile and the new one proposed here is to make the transition between the bursting layer and cool layer much smoother from the peak of the burst onwards. The buoyant $r$-mode is dependent on the jump in density at this boundary, with a sharper jump resulting in a larger rotating frame frequency. Accordingly, this new model for the initial condition will reduce the rotating frame frequency.

Similarly to PB05, the pre-burst state is found by solving Equation (\ref{eq:flux-NS-layer}) holding $F_r$ at a constant value set by the conditions at the crust. The temperature in the outer region of the layer approximately follows a radiative zero solution, $T \propto p^{0.24}$, and is insensitive to the boundary condition at the surface \citep{Schwarzschild58}. In the region where electron conduction dominates, deeper than $\sim 10^9$~\unity{}, the temperature slowly increases with depth, $T \propto p^{0.03}$. We test the effects of shallow heating by increasing the crustal flux to $10^{22}$~\unitF~for some calculations. This enhanced flux is appropriate for a NS accreting at $\dot{m} = 0.1$~$\mEdd$ with a crustal heating of $Q_{\mathrm{b}} = 1$~\MeVnuc~where $F_{\mathrm{crust}} = Q_{\mathrm{b}} \dot{m}$. The temperature profile for the heat pulse takes the shape of a Gaussian:
\begin{equation}
  \label{eq:heat-pulse}
  T_{\mathrm{hp}}(y) = T_{\mathrm{p}} \exp \left[ - \left( \frac{\log_{10} y - \log_{10} y_{\mathrm{c}}}{0.268} \right)^2 / 2 \right]
\end{equation}
where $T_{\mathrm{p}}$ is the peak of the heat pulse, $y_{\mathrm{c}}$ is the centre of the function, and the constant $0.268$ has been chosen such that $T_{\mathrm{hp}} (y_{\mathrm{c}} 10^{\pm 1}) / T_p \approx 10^{-3}$. The composition of each layer is taken to be the same as the PB05 model in order to isolate how a new initial condition affects mode frequencies.

During the rising phase of a real thermonuclear X-ray burst there are many violent, non-linear, and highly intricate processes occurring. Rapid, unstable nuclear burning converts light nuclei to heavier nuclei via many reaction chains \citep[for a review of the various burning processes that occur in the outer layers of NSs, see][]{Galloway17}. A burning front will spread from the ignition site across the star, rapidly engulfing the surface in flame \citep{Spitkovsky02}; this multidimensional process is not well understood and will be influenced by both the Coriolis force and the magnetic field configuration \citep{Cavecchi13,Cavecchi15,Cavecchi16}. Given these considerations, the `rising phase' in our model is not expected to be accurate and we consider our new ocean cooling model reasonable only after the peak of the burst where the temperature and flux at the surface of the star have stopped increasing.

We have avoided using the term `ignition depth' in our description of these models so far. For the purposes of calculating the frequency of the buoyant $r$-mode, the most important feature of the cooling model is $y_{\mathrm{t}}$, the location of the transition from the bursting layer to the cool layer. In PB05's model this location was considered to be the same as the ignition depth; for the heat pulse model they are not the same. Due to both the `rising phase' and the fact that the heat pulse has a width, the location $y_{\mathrm{c}}$ is not the same as $y_{\mathrm{t}}$. In order to achieve $y_{\mathrm{t}} = 3 \times 10^{8}$~\unity~at the time of the peak in flux we found using a shallower $y_c = 10^{8}$~\unity~to be appropriate.

\begin{table}
  \centering
  \caption{A summary of the parameters used in the new model for the initial conditions tested in this paper. All of these models include relativistic effects. Unless specified otherwise, all models assume a NS of mass $1.4$~\Msol{} and radius $10$~km.}
   \label{table:hp-models}
   \begin{tabular}{c r r r}
    Model & $y_{\mathrm{c}}$~(\unity) & $T_{\mathrm{p}}$~(GK) & $F_{\mathrm{c}}$~(\unitF) \\
    \hline
    PBEF & - & -   & $10^{22}$ \\
    HP11 & $10^8$ & 1.1 & $10^{22}$ \\
    HP10 & $10^8$ & 1.0 & $10^{22}$ \\
    HP09 & $10^8$ & 0.9 & $10^{22}$ \\
    F21  & $10^8$ & 1.1 & $10^{21}$ \\
    Y2   & $2 \times 10^8$ & 1.1 & $10^{22}$ \\
  \end{tabular}
\end{table}

Several sets of parameters are tested (see summary in Table \ref{table:hp-models}). We refer to cooling calculations with different sets of parameters as different `models' for cooling. Except in Section \ref{subsec:result-MR}, where the influence of the NS mass and radius on the mode is tested, all calculations include relativistic effects and use a NS of mass $1.4$~\Msol~and radius $10$~km, which results in a gravitational acceleration of $g = 2.4 \times 10^{14}$~\unitg~and volume correction factor of $\Vcorr = 1.31$. In order to compare the heat pulse model for the initial temperature profile to the previous work of PB05, a version of the PB05 initial condition will be examined with an enhanced flux emanating from the crust; this model will be referred to as PBEF. Three heat pulse models that resemble the temperature and density profiles in PBEF at the peak of the burst are inspected. The values of the peak of the heat pulse are $0.9$, $1.0$, and $1.1$~GK and will be referred to as HP09, HP10, and HP11 respectively. Each of these models uses the same enhanced flux emanating from the crust as PBEF, and $y_{\mathrm{c}} = 10^8$~\unity. The $T_{\mathrm{p}}$ and $y_{\mathrm{c}}$ parameters can be related to an average energy input to the layer and correspond to $(1.5, 1.9, 2.5) \times 10^{17}$~ergs~g$^{-1}$ for the models HP09, HP10, and HP11 respectively \citep[calculated using equation 2 of][]{Keek15}. Figure~\ref{fig:heat-pulse-background} shows the temperature profile $0.1$~s after the peak of the burst, and the flux emanating from outer boundary for the models HP09, HP10, HP11, and PBEF. At the peak of the burst, the temperature profile in the cool layer is the same for all four models because they have the same flux emanating from the crust, and heat from the bursting layer has not yet spread to these depths. The temperature gradient in the bursting layer is roughly the same for all models, with $T \propto y^{1/4}$, but each of the temperature profiles are offset from one another, leading to different density profiles in the bursting layer. The \bruntvaisala frequency and Lamb frequency, $\sqrt{\lambda} c_s / R$, at 0.1 s, 1.0 s, and 10 s after the peak of the burst are shown for the HP10 model in Figure~\ref{fig:heat-pulse-bv-lamb}.

Two more versions of the heat pulse model for the initial condition are examined. A model referred to as F21 uses a lower crustal flux of $10^{21}$~\unitF~and $y_{\mathrm{c}} = 10^{8}$~\unity~in order to demonstrate the effects of a lower flux from the crust. The final model, referred to as Y2, deposits the heat release by the burst at a deeper column depth of $y_{\mathrm{c}} = 2 \times 10^8$~\unity, but uses the same value for enhanced crustal flux. Both of these models use a peak temperature of $1.1$~GK, in order to best match the temperature and density profile of HP10 in the bursting layer. The temperature profiles of the models HP10, F21, and Y2 at $0.1$~s after the burst peak are shown in Figure~\ref{fig:heat-pulse-Fy}.

\subsection{Numerical method}

The solution of Equations (\ref{eq:T-evol-NS-layer}-\ref{eq:radius-NS-layer}) uses the method of lines whereby spatial derivatives are calculated using finite differencing, and the resulting set of points $T_i$ are the variables in an ODE in time which is solved using a stiff integrator. The grid is chosen to concentrate points about the transition from bursting to cool layers, and is uniform in $\sinh^{-1} \left[ \log (y / 3 \times 10^8 \text{\unity} ) \right]$. Gradients are calculated numerically upon this grid using the same finite difference scheme. At the inner boundary (at the crust) temperature is fixed, while the outer boundary holds the gradient of temperature fixed according to $\diff \log T/ \diff \log p \propto 1/4$, which is the case for a radiatively diffusive atmosphere. This is the same scheme used by \citet{Cumming04a}, \citet{Piro05b}, and \citet{Chambers18}. The layer always spans a column depth of $10^{5} - 10^{14}$~\unity~which covers the burning region of the ocean and the very outer portion of the crust (a distance of $\sim 10^{4}$ cm) and extends both shallower and deeper than the region considered by the mode calculation.

\section{Results}

Shallow-water wave analysis \citep{Pedlosky87,Berkhout08} can be used to estimate the frequency of the buoyant $r$-mode, and will be useful in explaining some of our results. Using a model in which the NS ocean is divided into two layers, the frequency of a surface wave is:
\begin{equation}
  \label{eq:freq-shallow}
  \sigma_r^2 \approx \Vcorr^{-1} g h k^2 \frac{\Delta \rho}{\rho} ,
\end{equation}
where $h = p / \rho g \Vcorr$ is the pressure scale height at the interface between the two layers, $k$ is the transverse wavenumber, and $\Delta \rho / \rho$ the jump in density between the two layers. The transverse wavenumber is related to the eigenvalue of Laplace's Tidal Equation as $k^2 = \lambda / R^2$, where $\lambda$ is given in Section \ref{subsec:laplace-equn}. The factor of $\Vcorr^{-1}$ is added based on arguments from \citet{Abramowicz02} of how to include gravitational redshift in the shallow-water equations. This expression for frequency should be valid at predicting some generic behaviours of the mode, such as how it scales with mass, radius, the density jump, and the transverse wavenumber. The upper layer will be considered as the hot, bursting portion of the ocean and the lower layer as the cool portion of the ocean.

\subsection{Purely relativistic effects on the mode frequency}
\label{sec:gr-effects-pb05}

\begin{figure}
  \centering
  \input{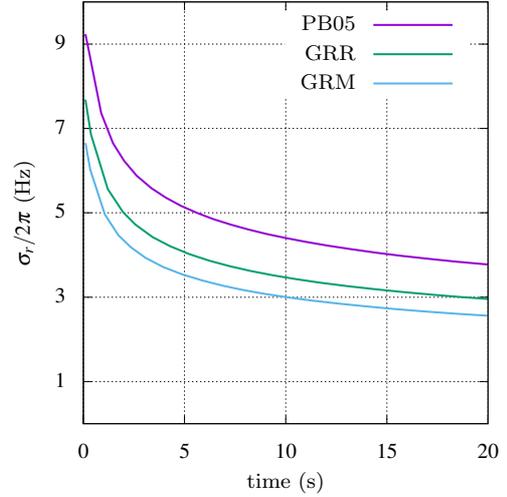}
  \caption{The rotating frame frequency of the $m=1$, $l=2$ buoyant $r$-mode in the cooling phase of the burst, demonstrating the influence of relativistic effects. The cooling model used is from PB05 for three cases: the Newtonian calculation (purple line) and two models that include the relativistic effects which match the gravitational acceleration of the Newtonian model and either the mass or the radius. 
  The green line shows the frequency change for model GRR that matches the radius of Newtonian model and the blue line shows the frequency change for model GRM that matches the mass of the Newtonian model. 
  More detailed information on the model parameters and the frequency change is given in Table \ref{table:PB05-params}.}
  \label{fig:PB05-compare}
\end{figure}

\begin{table*}
  \centering
  \caption{Detailed information from Figure~\ref{fig:PB05-compare} on the frequency of the mode at 15~s after the peak of the burst. The cooling models are summarised in Table~\ref{table:PB05-params}.
    $\Delta \sigma$ is the frequency drift; the change in frequency from measured from $0.1$~s until $15$~s after the peak of the burst.
    The third and fourth columns are the percentage change in rotating frame frequency and frequency drift when including relativistic effects for the models GRR and GRM when compared to PB05.
  }
  \label{table:PB05-freq}
  \begin{tabular}{l r r r}
    Model & $\Delta \sigma(15 \mathrm{~sec}) / 2 \pi$ (Hz) & $1 - \sigma_{r, i} / \sigma_{r, \mathrm{PB05}}$ & $1 - \Delta \sigma_{r, i} / \Delta \sigma_{r, \mathrm{PB05}}$ \\
    \hline
    PB05 & 5.2 & - & \\          
    GRR  & 4.6 & 22 \% & 11 \% \\
    GRM  & 4.0 & 32 \% & 23 \% \\
  \end{tabular}
\end{table*}

The relativistic effects on the buoyant $r$-mode are first tested using the models PB05, GRR, and GRM, which are based on the original work of PB05 using different parameters as described in Section \ref{subsec:init-cond}. Figure~\ref{fig:PB05-compare} shows the frequency of the buoyant $r$-mode as the burst evolves for these models. Table \ref{table:PB05-freq} lists detailed information about the parameters that set the model (mass, radius, and $\Vcorr$), the frequency drift at $15$ s after the peak of the burst, and the percentage change to the rotating frame frequency and the frequency drift that results from including relativistic effects.

Between the two models that include relativistic effects, changing the radius should have a bigger effect on the rotating frame frequency than changing mass (the rotating frame frequency from the model GRR should be smaller than that of GRM). The mass of the NS most significantly affects the mode through the gravitational acceleration at the surface, but this is the same for each model. Changing the radius affects the term $\lambda / R^2$ that appears in the surface wave estimate in Equation (\ref{eq:freq-shallow}). As a result, a larger radius should result in smaller rotating frame frequencies. The value for $\Vcorr$ is slightly different between models GRR and GRM due to the change in both mass and radius. According to the surface wave estimate in Equation (\ref{eq:freq-shallow}), a larger value of $\Vcorr$ should act to decrease frequency, which should further reduce the rotating frame frequencies calculated by GRM compared to GRR.

Based on the scaling behaviour of Equation (\ref{eq:freq-shallow}), considering only mass, radius, and $\Vcorr$, the rotating frame frequency of models GRR and GRM should relate to PB05 as:
\begin{align}
  \frac{\sigma_{r,\mathrm{GRR}}}{\sigma_{r,\mathrm{PB05}}} &\approx \Vcorr_{\mathrm{GRR}}^{-1/2} \left( \frac{M_{\mathrm{GRR}}}{M_{\mathrm{PB05}}} \right)^{1/2} \approx 0.8  , \\
    \frac{\sigma_{r,\mathrm{GRM}}}{\sigma_{r,\mathrm{PB05}}} &\approx \Vcorr_{\mathrm{GRM}}^{-1/2} \left( \frac{R_{\mathrm{GRM}}}{R_{\mathrm{PB05}}} \right)^{-2} \approx 0.7  .
\end{align}
Our results match these expectations. The rotating frame frequency for GRM is $\approx 30\%$ less than for PB05 throughout the course of the burst, while the frequency for GRR is $\approx 20 \%$ less than PB05.
The frequency drift exhibited by model GRM is $22\%$ smaller than PB05 while the model GRR exhibits frequency drift smaller by $10\%$ at the same time. This test shows that including gravitational redshift and frame-dragging can significantly alter the buoyant $r$-mode model.

\subsection{Effects from the heat pulse initial condition}
\label{subsec:res-hp-effects}

\begin{figure}
  \centering
  \input{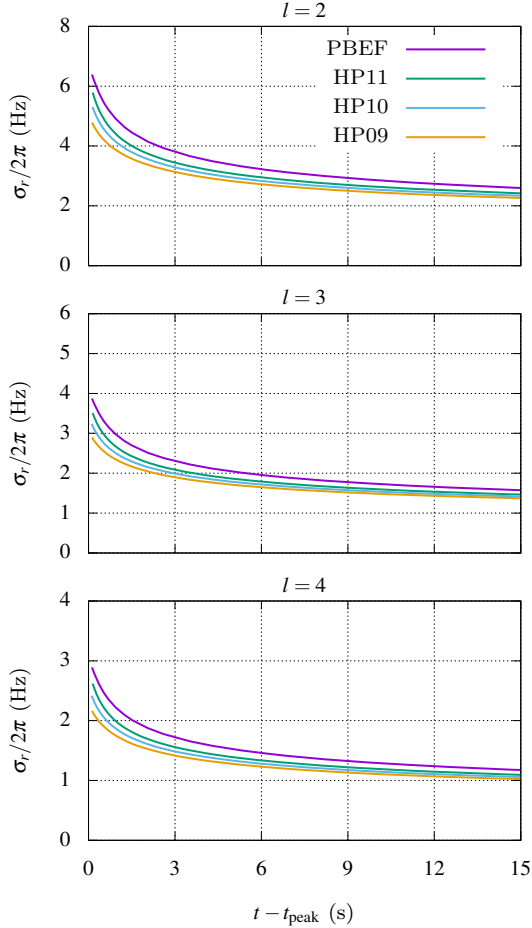}
  \caption{
    The rotating frame frequency of the $m=1$, $l=2, 3$, and $4$ buoyant $r$-modes. All of these calculations assumed a NS of mass $1.4$~\Msol~and radius $10$~km, and include relativistic effects.
    The purple lines are for the PBEF cooling model (PB05's calculation including an enhanced crustal flux). The green, blue and  orange lines are for the Gaussian model with peak temperatures $1.1$, $1.0$, and $0.9$~GK respectively (models HP11, HP10, and HP09). These cooling models are shown in Figure~\ref{fig:heat-pulse-background}.}
  \label{fig:heat-pulse-freq}
\end{figure}

\begin{figure}
  \centering
  \input{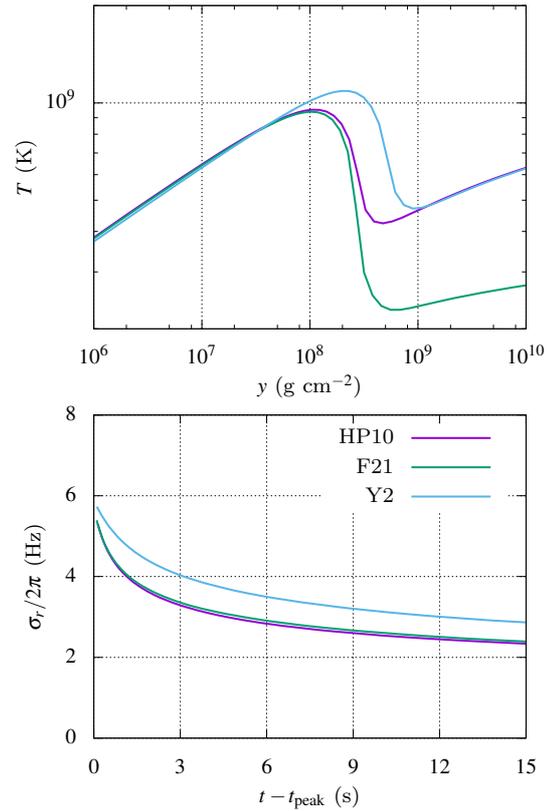}
  \caption{The upper panel shows the temperature profiles $0.1$~s after the peak of the burst for three models that investigate the influence of crustal heating and location at which nuclear burning energy is deposited on the $l=2$ buoyant $r$-mode. The parameters of each model are given in Table \ref{table:hp-models}.
    The difference between model HP10 and F21 is the value of the flux emanating from the crust: $10^{22}$~\unitF~and $10^{21}$~\unitF~were used in the case of HP10 and F21 respectively. The difference between the model HP10 and Y2 is the value of $y_{\mathrm{c}}$: HP10 uses a value of $10^{8}$~\unity~while Y2 uses a value of $2 \times 10^{8}$~\unity.
The lower panel shows the rotating frame frequency of the $m=1$, $l=2$ buoyant $r$-mode in the aftermath of a burst for the same models.}
  \label{fig:heat-pulse-Fy}
\end{figure}

We now inspect how the frequency of buoyant $r$-modes changes depending on the initial conditions for the temperature profile. We use the cooling model HP10, based on the heat pulse initial condition, to compare to PBEF, based on PB05 model for initial conditions as described in Section \ref{subsec:init-cond}. HP10 most closely resembles the temperature and density profile in the bursting layer of PBEF.
From the estimate for the surface wave in Equation (\ref{eq:freq-shallow}), the important factors for explaining the difference in the rotating frame frequency of the buoyant $r$-mode between each model for cooling are the pressure scale height and density jump between the bursting and cool layers. The volume correction factor, gravitational acceleration, and mode wavevector are the same for each of these models.

The pressure scale height at the transition from bursting to cool layers is approximately the same for both of the cooling models, with $h \approx 1.8 \times 10^{2}$~cm for PBEF and $h \approx 1.4 \times 10^{2}$~cm for HP10. The density jump between the bursting and cool layers may be estimated from the peak in density gradient at the transition between the two layers, $\diff \log \rho / \diff \log p$. At $0.1$~s after the peak of the burst this quantity is $1.75$ for the model PBEF and $1.25$ for HP10. These effects explain the higher rotating frame frequency of the buoyant $r$-mode for the PBEF model compared to the HP10 model. At $0.1$~s after the burst peak, the mode frequency calculated using the PBEF model for cooling is $6.4$~Hz, while for the HP10 mode it is $5.3$~Hz. 

The most important factor for determining the drift in frequency of the buoyant $r$-mode is the magnitude of the change in the density jump between the bursting and cool layers between the start and end of the burst.
At $15$~s after the peak, the density gradient at the transition between the two layers is $0.75$ for both This explains why the frequency drift of the PBEF model, $3.9$~Hz, is larger than that of the HP10 model, $3$~Hz.

\subsection{Dependence on the heat pulse parameters}
\label{subsec:result-hp}

Increasing the peak temperature in the heat pulse models (increasing the $T_{\mathrm{p}}$ parameter) increases the jump in density across the transition from bursting to cool layers. This increases the rotating frame frequency, as can be seen from Figure~\ref{fig:heat-pulse-freq}, which shows a spread of frequencies of $4.8 - 5.8$~Hz at $0.1$~s between the different models and $2.3 - 2.4$~Hz at $15$~s. The frequency drift also increases with peak temperature; the models tested here show a spread of frequency drifts of between $2.5 - 3.4$~Hz.

Depositing the heat released from nuclear burning at a deeper column depth (using a deeper $y_{\mathrm{c}}$) changes the pressure scale height at the transition from bursting to cool layers. For the Y2 model this occurs at $2 \times 10^{2}$~cm which increases the rotating frame frequency of the buoyant $r$-mode. The more significant effect of the deeper $y_{\mathrm{c}}$ is on the cooling timescale of the burst; the longer cooling timescale reduces the frequency drift exhibited by the buoyant $r$-mode calculated on this background. This can be seen from Figure~\ref{fig:heat-pulse-Fy}. At $15$~s after the burst, the buoyant $r$-mode calculated on the Y2 cooling model shows a frequency drift of $2$~Hz and rotating frame frequency $0.4$~Hz higher than that of the mode calculated for the HP10 model. However, the depth at which the transition occurs is set by the conditions for unstable thermonuclear burning to trigger. Using a greater value for $y_{\mathrm{c}}$ in an effort to curtail drifts may result in an unphysical ignition site.

The enhanced flux emanating from the crust will increase the temperature of the cool layer, which will have two effects. The first is to increase the jump in density across the transition from bursting to cool layer, and the second effect is on the rate at which the density jump changes with time. However, both of these effects are small. The cool ocean is dominated by degenerate electrons and so an increase in temperature of the cool layer should have a small effect on the density jump. Nevertheless, this jump should always be greater for a model with a cooler ocean and therefore a buoyant $r$-mode that exists upon the F21 cooling model should always have a higher rotating frame frequency than the HP10 model. For a cool layer dominated by degenerate electrons, the conductivity and heat capacity of the fluid both depend linearly on temperature. This means that the thermal diffusivity, $K / \rho c_P$, is weakly dependent on flux, therefore the rate at which heat spreads to the cool layer should be approximately the same. The rotating frame frequency of the mode calculated using the model HP10 is always smaller than that of F21. The frequency drift of the buoyant $r$-mode $15$~s after the peak for the HP10 model is greater than the F21 model by $0.1$~Hz.

The mode is sensitive to the location of the outer boundary, which was chosen at a depth such that the mode timescale is equal to the thermal timescale. Changing this depth has a small effect on the mode frequency because the energy of the mode is concentrated at the location of the jump in density which is always shallower than $10^8$~\unity{} in column depth; the frequency of the mode increases by $< 0.5$~Hz when the location of the outer boundary is made shallower, at $10^6$~\unity{}. The depth of the outer boundary cannot be made shallower than $10^6$~\unity{} since the adiabatic condition would no longer be valid.

\subsection{Higher $l$ buoyant $r$-modes}
\label{subsec:result-high-l}

The difference in rotating frame frequency between the $l=2$, $l=3$, and $l=4$ buoyant $r$-modes can be explained from their dependence on the transverse wavenumber and thus $\lambda$. The surface wave estimate, Equation (\ref{eq:freq-shallow}), predicts that the rotating frame frequency should scale as $\sigma_r \propto \sqrt{\lambda}$. For the modes investigated here we have  $\lambda_{2} = 0.11$, $\lambda_{3} = 4.1 \times 10^{-2}$ and $\lambda_{4} = 2.2 \times 10^{-2}$ for the $l=2$, $l=3$, and $l=4$ buoyant $r$-modes respectively.
Given these considerations we should expect the rotating frame frequency of the buoyant $r$-modes to be related as
\begin{align}
\frac{\sigma_{r,3}}{\sigma_{r,2}} &= \left( \frac{\lambda_3}{\lambda_2} \right)^{1/2} \approx 0.6 , \\
\frac{\sigma_{r,4}}{\sigma_{r,2}} &= \left( \frac{\lambda_4}{\lambda_2} \right)^{1/2} \approx 0.4 .
\end{align}
These estimates match our results. For the HP10 model, the rotating frame frequency at $15$~s after the peak of the burst is $2.4$~Hz, $1.4$~Hz, and $1.0$~Hz for the $l=2$, $l=3$, and $l=4$ buoyant $r$-modes respectively.

\subsection{Mass and radius effects}
\label{subsec:result-MR}

\begin{figure}
  \centering
  \input{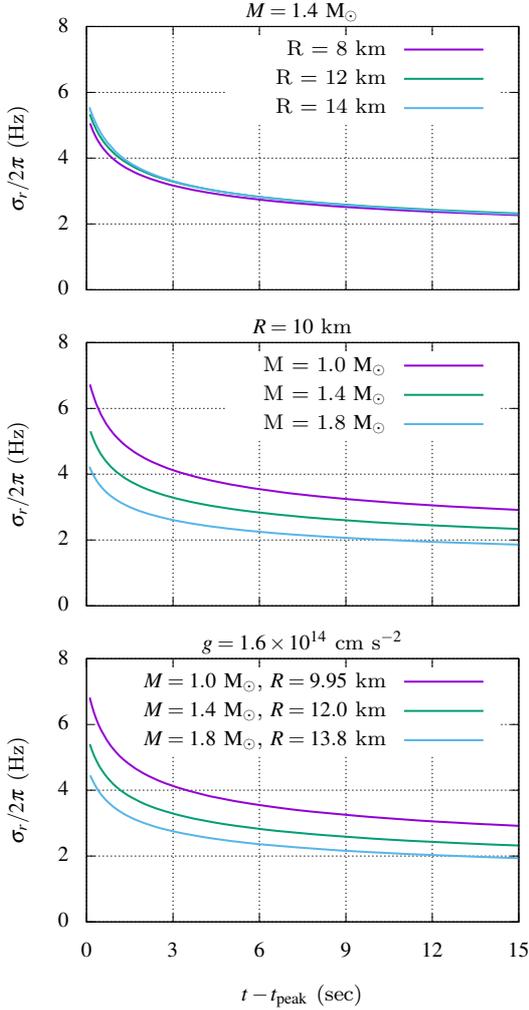}
  \caption{The frequency of the $l=2$ buoyant $r$-mode for the HP10 model of the cooling ocean for a range of masses and radii. The panels show calculations with a common mass (top), radius (middle), or gravitational acceleration at the surface of the NS (bottom). Only calculations with fixed gravitational acceleration obey the scaling relations in Equation (\ref{eq:freq-shallow-MR}).}
    \label{fig:freq-MR}
\end{figure}

Estimates for the radius of a 1.4~\Msol{} NS given in \citet{Hebeler13} predict a range of radii between $8.8 - 13.9$~km. This begs the question of how the frequency of the buoyant $r$-mode depends on mass and radius. In this section, the effect of changing the mass and radius of the NS on the buoyant $r$-mode frequency is outlined. The first result of changing these parameters is to alter the wavevector $k^2 = \lambda / R^2$. The second is to alter the ocean cooling model upon which modes are solved, since this depends on the gravitational acceleration at the surface.

Several calculations of the rotating frame frequency of the $l=2$ buoyant $r$-mode are performed, using the HP10 model for the cooling ocean for a range of masses and radii. The results are shown in Figure \ref{fig:freq-MR}. Masses range from $1.0 - 1.8$~\Msol, and radii from $8-14$~km. Solutions are grouped into panels with a common mass, radius, or gravitational acceleration at the surface of the NS. Equation (\ref{eq:freq-shallow}) predicts the rotating frame frequency of the buoyant $r$-mode, and helped to explain how frequency scales with $\lambda$ in Section \ref{subsec:result-high-l}. We now compare this estimate to the full calculation of the buoyant $r$-mode that includes the radial structure and cooling in the aftermath of the burst.

With parameters appropriate for the HP10 model, the frequency estimate becomes:
\begin{align}
  \label{eq:freq-shallow-MR}
  \nonumber
  \frac{\sigma_r}{2 \pi}
  \approx {} &
  8 \text{ Hz} \,
  \left( \frac{\Vcorr}{1.3} \right)^{-1/2}
  \left( \frac{M}{1.4 \text{ \Msol}} \right)^{1/2}
  \left( \frac{R}{10 \text{ km}} \right)^{-2}
  \left( \frac{\lambda}{0.11} \right)^{1/2} 
  \\
  & \times \left( \frac{h}{1.4 \times 10^{2} \text{ cm}} \right)^{1/2}
  \left( \frac{\Delta \rho}{\rho} \right)^{1/2}
  .
\end{align}
Note that the background cooling model dictates $h$ and $\Delta \rho$; the background, in turn, depends on the mass and radius through the gravitational acceleration and volume correction factor (except for a small term in Equation \ref{eq:T-evol-NS-layer}). Since $h$ and $\Delta \rho$ follow from the background, we expect their dependence on mass and radius to manifest only through the gravitational acceleration and volume correction factor.

The top panel of Figure \ref{fig:freq-MR} shows no change in rotating frame frequency when changing radius, even though there should be a strong $R^{-2}$ dependence according to Equation (\ref{eq:freq-shallow-MR}). The middle panel shows a much stronger dependence on the mass, however the change in rotating frame frequency with mass follows the opposite trend to expectations. Equation (\ref{eq:freq-shallow-MR}) predicts a smaller rotating frame frequency for smaller masses, while the calculation shows an increasing rotating frame frequency with smaller masses. 

Changing mass and radius while holding the gravitational acceleration fixed does match the expectations of Equation (\ref{eq:freq-shallow-MR}). The rotating frame frequencies at $15$~s after the peak of the burst are  $2.9$~Hz, $2.3$~Hz, and $1.9$~Hz for the modes that use a NS mass of $1$~\Msol, $1.4$~\Msol, and $1.8$~\Msol{} respectively. These frequencies of two calculations with differing mass, radius and buoyant $r$-mode rotating frame frequency can be related using the formula:
\begin{equation}
\frac{\sigma_{r,a}}{\sigma_{r,b}} \approx  \left( \frac{\Vcorr_a}{\Vcorr_b} \right)^{-1/2} \left( \frac{M_a}{M_b} \right)^{1/2} \left( \frac{R_a}{R_b} \right)^{-2}.
\end{equation}
These results show that the estimate in Equation (\ref{eq:freq-shallow-MR}) is only valid when holding gravitational acceleration fixed (as was done in Section \ref{sec:gr-effects-pb05}) because of the effect on the cooling model. This work considered column depth to be the most important variable used to define: the size of the layer; the location that nuclear energy is deposited; and the region in which the mode exists. Changing the gravitational acceleration of the cooling calculation altered the pressure and density at these locations, and hence changed the frequency of the mode. 

Next, we estimate the smallest viable rotating frame frequency and frequency drift that the buoyant $r$-mode model can achieve using the heat pulse model for the initial condition. The model HP09 was used as it gave the smallest values for rotating frame frequency and frequency drift. A NS of mass $1.8$~\Msol{} and radius $14$~km was used to further reduce the rotating frame frequency. At $15$~s after the peak of the burst, the rotating frame frequency of the mode calculated using this model is $1.7$~Hz, and the frequency drift is also $1.7$~Hz. Using a smaller value for $y_{\mathrm{c}}$ should also reduce the rotating frame frequency, but may also result in a larger frequency drift.

\subsection{Changes to the surface pattern}
\label{subsec:surface-pattern}

\begin{figure}
  \centering
  \input{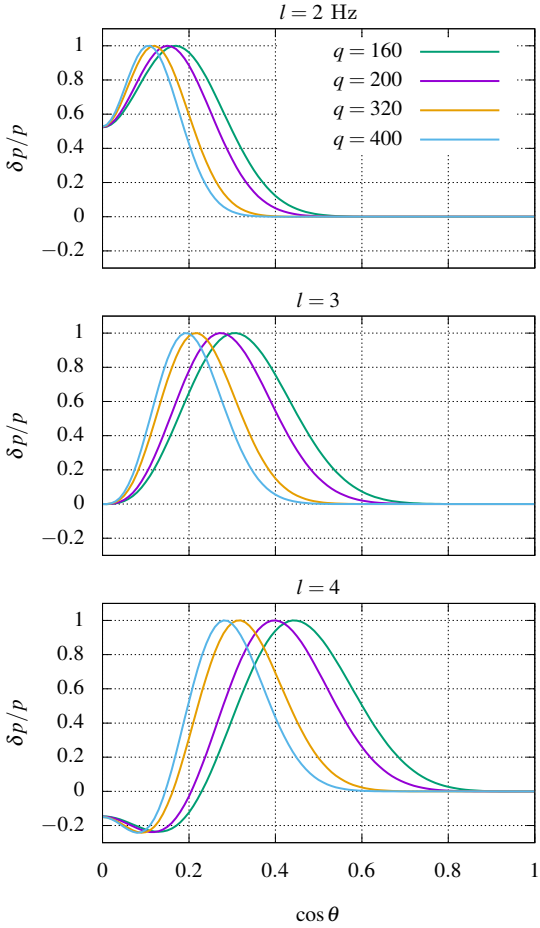}
  \caption{The co-latitudinal variation of the $l=2, 3$ and $4$ buoyant $r$-mode with rotating frame frequency of $3$~Hz, for several values of rotation rate. Each panel plots a different value of $l$, within each panel is a different value of the spin parameter, $q$, that is affected by frame-dragging. The relativistic and Newtonian values can be related as: $q = q_{\mathrm{N}} ( 1 - \omega / \Omega )$.
    The purple (blue) line uses the Newtonian definition of the spin parameter to calculate the eigenfunction for a NS spinning at 300~Hz (600~Hz) while the green (orange) line includes the factor of $( 1 - \omega / \Omega ) = 0.8$ for a NS rotating at 300~Hz (600~Hz).}
  \label{fig:eigfunc-pair}
\end{figure}

Another important consideration when including relativistic effects is the change to the surface pattern of the buoyant $r$-mode. The co-latitudinal dependence of the mode is found from the eigenfunction of Laplace's Tidal Equation which depends on the spin parameter, $q$. Frame-dragging shrinks $q$ by a factor $(1 - \omega / \Omega)$ compared to the Newtonian value (see Equation \ref{eq:spin-param}). For buoyant $r$-modes, although the eigenvalue of Laplace's Tidal equation is constant with $q$, the eigenfunction will change with this parameter, as discussed in Section \ref{subsec:laplace-equn}.

Frame-dragging is evaluated at the surface of the NS and depends on the NS's moment of inertia. We use the results of \citet{Baubock13} that numerically calculated the moment of inertia of a NS for a range of masses, radii, and core equations of state in a space-time described by the Hartle-Thorne metric. They provide empirical fits for the dependence of the moment of inertia on compactness, $GM / c^2 R$, using a polynomial. For a NS of mass $1.4$~\Msol~and radius $10$~km, the change in the spin parameter when including frame-dragging is: $(1 - \omega / \Omega) \approx 0.8$.

Assuming a buoyant $r$-mode with rotating frame frequency $\sigma_r / 2 \pi = 3$ Hz, which is consistent with our results, and a NS spin frequency of $300$~Hz and $600$~Hz results in Newtonian spin parameter $q_{N} = 200$ and $q_{N} = 600$ respectively. When frame-dragging is included this spin parameter shrinks to $q = 160$ and $q = 320$. This decrease in the spin parameter acts to reduce the equatorial confinement of the mode. The change in the co-latitudinal dependence of the surface pattern when including the effect of frame-dragging is plotted in Figure~\ref{fig:eigfunc-pair} for the $l=2$, $l=3$, and $l=4$ buoyant $r$-mode for NSs spinning at $300$~Hz and $600$~Hz.

There is a marginal difference in the location of the peak in pressure perturbation for the $m=1$, $l=2$ buoyant $r$-mode between the Newtonian and relativistic calculations, and progressively greater differences for the $l=3$, and $l=4$ buoyant $r$-mode. We have assumed that the mode is adiabatic, making the temperature perturbation proportional to the pressure perturbation, hence these results should indicate how relativistic effects influence the temperature pattern.

\section{Discussion}

In this paper we have investigated the frequency of buoyant $r$-modes in the aftermath of a thermonuclear X-ray burst. Previous calculations of this mode have found rotating frame frequencies and frequency drifts that are too large to be consistent with observations. Our new calculations included relativistic effects, which act to reduce both the rotating frame frequency and the frequency drift exhibited by the mode. A simple cooling model, that removes large discontinuities in density in the ocean compared to previous work, was found to further reduce both the rotating frame frequency and frequency drift. For a NS of mass $1.4$~\Msol{} and radius $10$~km, the rotating frame frequency of the buoyant $r$-mode $15$~s after the peak of the burst was $2.3 - 2.4$~Hz, and the frequency drift (from the burst peak) was in the range $2.4 - 3.3$~Hz, depending on the quantity of energy deposited in the ocean by nuclear burning. Altering the mass and radius, within the range of values considered feasible by current dense matter equation of state models, could reduce the rotating frame frequency to as low as $1.7$~Hz, and frequency drift to $1.7$~Hz (see Section \ref{subsec:result-MR}).

\subsection{Mass, radius, and gravitational redshift}

Gravitational redshift and frame-dragging have a marked effect on buoyant $r$-modes in the oceans of NSs. These relativistic effects decrease the rotating frame frequency of the mode by up to $30 \%$, which is larger than the estimate of $15 - 20 \%$ reported by MA04. This difference is most likely due to the fact that the frequency calculations in our work also account for the radial structure of the mode. Relativistic effects can also reduce the frequency drifts, by up to $20 \%$.

The mass and radius of the NS affect mode frequencies in two ways: the first is a direct effect on the properties of the mode through the wavevector that scales with $1 / R$; the second is via changes to the ocean cooling model. In the ocean cooling model, the gravitational acceleration at the surface determines the pressure scale height and the jump in density at the column depth where the burst ignites. The influence of these factors on the mode can be estimated from Equation (\ref{eq:freq-shallow}). However, as shown in Section \ref{subsec:result-MR}, applying these scalings to mode frequencies should be performed with caution. The buoyant $r$-mode only obeys the scaling relation in Equation (\ref{eq:freq-shallow}) when comparing models with the same value of gravitational acceleration, due to the fact this quantity affects the cooling background upon which modes are computed.

\subsection{A new cooling initial condition}

The new model for the initial conditions reduces the rotating frame frequency when compared to the original model examined by PB05. Including general relativistic effects and using a NS of mass $1.4$~\Msol{} and radius $10$~km, the model for ocean cooling using the PB05 parametrisation of the initial condition (which had a sharp discontinuity in density at the ignition depth) results in a buoyant $r$-mode rotating frame frequency of $2.7$~Hz and frequency drift of $3.9$~Hz at a time $15$~s after the peak of the burst. The heat pulse cooling model (which smoothed the discontinuity in density at the ignition depth) that most closely resembled the temperature and density profile of this calculation resulted in a rotating frame frequency of $2.3$~Hz and frequency drift of $3$~Hz at the same time. This reduction in both the rotating frame frequency and frequency drift for the heat pulse model is explained by the smaller jump in density between the bursting layer and cool layer.

Buoyant $r$-modes showed a marginal change in the rotating frame frequency and frequency drift when the crustal flux is changed; approximately a $0.1$~Hz change at $15$~s after the peak. This shows that the buoyant $r$-mode has little dependence on a change in base flux when holding all other parameters fixed. More important to the frequency were the quantity of and depth at which energy from the burst were deposited. The base flux will affect the cooling model and burst conditions, therefore the mode frequency will be altered through these effects.

\subsection{Comparison with observed TBO frequencies and drifts}

The frequency drifts from TBO sources that do not exhibit accretion-powered X-ray pulsations can range from $1 - 3$~Hz in the tail of a burst \citep{Muno02a,Galloway08,Bilous18}. The model with the smallest heat released from nuclear burning exhibited a $l=2$ buoyant $r$-mode with the smallest frequency drift of $2.3$~Hz, which is still too large for the lower limits of the observed range. By using another set of parameters for mass and radius, the drift in frequency can be reduced to as low as $1.7$~Hz. A model that deposits the energy released from the burst at a shallower depth would further frequency drift.

Sources that intermittently show accretion-powered pulsations provide an independent measure on the difference between the TBO frequency and the spin frequency.  These sources show a spin $1$~Hz higher than the TBO frequency observed during an X-ray burst (see \citealt{Casella08} for detail on Aql~X-1, and for HETE~J1900.1-2455 see \citealt{Watts09}). These observations provide a tight constraint on the rotating frame frequency of a buoyant $r$-mode, which must be as low as $1$~Hz. This constraint is not satisfied for the relativistic $l=2$ buoyant $r$-mode for any of the ocean cooling models tested here, which do not show rotating frame frequencies smaller than $1.7$~Hz. 

The observation of accretion-powered pulsations indicates that the source has a strong magnetic field \citep[see][]{Poutanen06}. The presence of a magnetic field may significantly alter the mode (for instance, see calculations by \cite{Heng09} and \cite{Marquez17}). The nature of TBOs observed from such sources, however, is quite different from those with no accretion-powered pulsations observed. Our results should only apply to sources with no pulsations present during the period in which TBOs are observed.

\subsection{Higher $l$ modes as a potential TBO mechanism}

The $l=4$ buoyant $r$-mode shows a smaller rotating frame frequency than the $l=2$ mode at $\sim 1$~Hz, and a smaller frequency drift of $\sim 1$~Hz. The parameters of the cooling model can be changed to alter the drifts enough to reach $3$~Hz, which is the upper limit of the observed range. These modes might be better candidates to explain TBOs, since they show rotating frame frequencies and frequency drifts low enough to be consistent with TBO observations. This argument does not consider the pulsed amplitude that the mode should exhibit which should be smaller for higher $l$ modes \citep{Lee05}.

Another consideration is the potential preferential excitation of a $l=3$ or $l=4$ buoyant $r$-mode over a $l=2$ mode. Work exploring the relationship between accretion rate, rotation rate and ignition latitude has been performed by \citet{Cooper07d} and \citet{Cavecchi17} and showed that, depending on the interplay between these conditions, bursts may be preferentially ignite at higher latitudes. Ignition latitude should have an influence on what mode is excited, so it does not seem unreasonable that bursts ignited at higher latitude might preferentially excite modes more concentrated at higher latitudes. However, these studies also indicate that off-equator ignition is rare and only viable for rotation rates $\gtrsim 600$~Hz, which does not apply to all TBO sources. A strong magnetic field may affect the ignition latitude also \citep{Cavecchi16}.

Another explanation might come from calculations of flame spread across the NS ocean in the aftermath of a burst. Recent simulations of flame spread during the rising phase of a burst by \citet{Cavecchi19} have shown that large scale vortices can be excited during the rising phase of the burst (of the type that resemble these buoyant $r$-modes). Vortices appear in the aftermath of the explosion as a result of the baroclinic instability; a misalignment of pressure and density gradients in a fluid (see \citealt{Gill82}, \citealt{Pedlosky87}, and in the context of NS oceans see \citealt{Cumming00}). The fact that these vortices can appear at high latitudes would seem to indicate that higher $l$ modes are potentially excitable in the tail of a burst. A calculation of the excitation mechanism of these modes needs to be performed to answer this issue.

Studies performed by \citet{Strohmayer96a}, \citet{Piro04}, and \citet{Narayan07} inspected how an $\epsilon$-mechanism might excite a buoyant $r$-mode. In particular, \citet{Piro04} found that modes with a low number of radial nodes ($n=2$ as studied in the current work) were preferentially excited over modes with a higher number of radial nodes. \citet{Piro04}, however, assumed spherical symmetry of the background; we anticipate that multidimensional effects previously mentioned (as well as possible differential rotation, see \cite{Cumming05a}) will be highly relevant when evaluating whether modes with different latitudinal number are preferentially excited.

\subsection{The surface pattern and pulsed amplitude}

Previous calculations of the pulsed amplitudes of buoyant $r$-modes by \citet{Lee05}, and \citet{Heyl05} predicted smaller amplitudes than those of TBO observations. The observed pulsed amplitudes vary from burst to burst, being as low as $5 \%$ (the detectability limit) and as high as $20 \%$ for the sources 4U~1728-34, 4U~1702-429, and 4U~1636-536 \citep{Muno02b,Ootes17}. \citet{Piro06} calculated the energy dependence of pulsation amplitudes, finding favourable agreement between predictions and observations. The results from our calculation show that relativistic effects change the surface pattern of the mode by inhibiting equatorial trapping (see Figure \ref{fig:eigfunc-pair}), but this effect is marginal.

Two key results from \citet{Heyl05} are that decreasing $q$ while holding $\Omega$ fixed increases the pulsed amplitude, and increasing $\Omega$ while holding $q$ fixed increases the pulsed amplitude but to a lesser extent. This fact seems promising given that the extra factor of $1 - \omega / \Omega$ reduces $q$ for the same $\Omega$. However, the decrease in $q$ is much too small to make a large difference in amplitude. From the work of \citet{Heyl05}, decreasing $q$ from 600 to 300 increases the pulsed fraction from $\sim 0.03$ to $\sim 0.04$ which is a much greater change in $q$ than reasonable here, where $1 - \omega / \Omega \approx 0.8$ (see Section \ref{subsec:surface-pattern}). It would seem that in order to increase the pulsed amplitudes to observable levels, including relativistic effects is not enough.

The pulsed fraction of the $l=3$ buoyant $r$-mode has been studied in \citet{Lee05} and was found to be smaller than the $l=2$ mode by $\sim 50 \%$. As to whether the $l=4$ buoyant $r$-mode should exhibit a greater pulsed amplitude than the $l=2$ mode, this has not been studied. One overarching problem with amplitude calculations is that the normalisation of the buoyant $r$-mode is not predicted from the linear theory used to calculate the the angular dependence of the mode. This problem was discussed in Section \ref{sec:perterbation-equations} but left out of the calculation as it had no influence on frequency or frequency drift.

\subsection{Influence of composition and ongoing nuclear burning}
\label{subsec:dis-nucback}

A previous study by \citet{Chambers19} calculated the frequency evolution of buoyant $r$-modes using a model for the cooling ocean in the aftermath of a burst from \citet{Keek17}. This cooling model included a large nuclear reaction network that calculated nuclear burning throughout the course of the burst and accounted for changing composition. \citet{Chambers19} showed that the frequency drift of the buoyant $r$-mode was $< 4$~Hz as a result of the slow temperature change at the ignition site. Our results in the current work show that including gravitational redshift with this model would reduce the frequency drift further.

\citet{Chambers19} also found a large rotating frame frequency of the buoyant $r$-mode, which was $5 - 8$~Hz and does not match the expectation based on observations of TBOs. We note two properties of the cooling model used in \citet{Chambers19} that contribute to the rotating frame frequency. The first is the smooth temperature profile which should act to reduce rotating frame frequencies (as shown in Section \ref{subsec:res-hp-effects}). The second is a light composition in the bursting layer. The compositional effect lowers the jump in density at the transition from the bursting to the cool layer, which, in competition with the smooth temperature, raises the rotating frame frequency of the mode. The results presented in the current work show that gravitational redshift can reduce rotating frame frequencies by as much as $30 \%$ compared to a Newtonian model, which was used in \citet{Chambers19}. However, this will likely not be enough to match the constraints of the intermittent accretion-powered pulsars that require a rotating frame frequency of $1$~Hz. Indeed, the heat pulse model for ocean cooling, examined in this work, used a heavier composition in the bursting layer of pure $^{40}$Ca and included gravitational redshift, and still is not in the range of observed TBOs.

The cooling model of \citet{Keek17} used a large value for the base flux at $Q_{\mathrm{b}} = 3$~\MeVnuc. This parameter has little direct effect on the buoyant $r$-mode frequency, as shown in Section \ref{subsec:result-hp}, altering the mode primarily through the influence on the background state and burst conditions.

\section{Conclusions}

Although the buoyant $r$-mode model has many attractive features as a potential mechanism to explain TBOs, it appeared to suffer from two problems in particular. First, the drift in frequency was too great compared to the observed drifts in the tail of bursts. Secondly, the rotating frame frequency of the mode was greater than the offset observed between the asymptotic frequency of TBOs and the spin frequency of the NS (where known independently). Some key pieces of physics were, however, not taken into account in the first calculations of the buoyant $r$-mode frequency. The interplay between two of these effects, ongoing nuclear burning and gravitational redshift, is now discussed.

\citet{Chambers19} showed that the frequency drift of the buoyant $r$-mode can be curtailed by including nuclear burning throughout the course of the burst. This resulted in a frequency drift $< 2$~Hz over 15~s in some cases. In this current work, building on the calculation of MA04, we have shown that including gravitational redshift and frame-dragging reduces frequency drifts by $\sim 20 \%$ compared to an equivalent Newtonian model with the same gravitational acceleration at the surface of the NS. It would seem reasonable to expect that a calculation including both of these effects might exhibit frequency drift over the observed range of $1-3$~Hz.

The cooling model used in \citet{Chambers19} also showed that the rotating frame frequency of the buoyant $r$-mode increased to $\sim 5$~Hz as a result of the lighter ocean composition present in that model. Gravitational redshift may reduce this frequency by $\sim 30 \%$, but it is still too large to explain the difference between the TBO and spin frequency, for the specific cooling model studied in \citet{Keek17}. The current work indicates that higher $l$ modes may be a potential avenue to resolve this issue.

\citet{Chambers19} and the current work demonstrate that including previously neglected physics may return the buoyant $r$-mode model to the status of being a viable candidate to explain TBOs. The buoyant $r$-mode model now needs to be tested including gravitational redshift, using a variety of burst models that account for detailed burning, and a variety of NS properties.

\section*{Acknowledgements}

The authors acknowledge support from ERC Starting grant No. 639217 CSINEUTRONSTAR (PI: Watts). This work benefited from discussions at the BERN18 Workshop supported by the National Science Foundation under grant No. PHY-1430152 (JINA Center for the Evolution of the Elements). We thank Andrew Cumming for his suggestion that ignition condition might preferentially excite the $l=4$ mode. We thank the Ioffe Institute for making their equation of state routines publicly available.



\bibliographystyle{mnras}
% \bibliography{rmodes,rmodes-extra}
% for arXiv submission

%%%%%%%%%%%%%%%%%%%%%%%%%%%%%%%%%%%%%%%%%%%%%%%%%%
%%%%%%%%%%%%%%%%% APPENDICES %%%%%%%%%%%%%%%%%%%%%

% \appendix

%%%%%%%%%%%%%%%%%%%%%%%%%%%%%%%%%%%%%%%%%%%%%%%%%%

% Don't change these lines
\bsp	% typesetting comment
\label{lastpage}
\end{document}